\documentclass[12pt,a4paper]{article}
\usepackage{amsmath}
\usepackage{graphicx}
\usepackage{times}
\begin{document}
\begin{titlepage}
\title{Secondary dips and the asymptotics}
\author{ S.M. Troshin, N.E. Tyurin\\[1ex]
\small  \it SRC IHEP of NRC ``Kurchatov Institute''\\
\small  \it Protvino, 142281, Russian Federation\\
}
\normalsize
\date{}
\maketitle

\begin{abstract}
We point out how to detect experimentally the energy region where   asymptotics starts to manifest itself in hadron scattering relating appearance of the secondary dips in the differential cross-section of elastic scattering $d\sigma/dt $  with  beginning of the asymptotic energy region. The consideration relies on the differential characteristics. The framework of the impact parameter picture of proton--proton interactions is being used.  
\end{abstract}
\end{titlepage}
\setcounter{page}{2}
We discuss  a qualitative experimental signature of the  asymptotics   in $pp$--scattering, proceeding from  consideration of the total cross-section of proton interactions. The latter is determined by the imaginary part  of the elastic scattering amplitude $\mbox{Im}f(s,b)$ integrated over impact parameter $b$. 

The correlations between the particular qualitative features of the amplitude in the impact parameter space and the structure of the differential cross--section $d\sigma/dt$ were  described in \cite{halzen} long time ago. In particular, it was shown that developing dip in $d\sigma/dt$ in the region of $-t=1.4$ $(GeV/c)^2$ should be associated with increasing role of unitarity with the energy growth, i.e with the absorption at low impact parameter values. In \cite{laslo} recent discussion can be found in connection with the TOTEM data obtained at the LHC \cite{totem}.  In general, the differential cross--section $d\sigma/dt$ at large values of $-t$ is most sensitive to the region of low impact parameter values.  Such correlation is exploited here to point out the experimental signature of the asymptotic energy region on the grounds of the   differential cross--section $d\sigma/dt$ behavior in the region of large transferred momenta.

Following Chew and Frautchi \cite{chew} we relate start of the  asymptotic energy  region with approach to the maximal strength of strong interactions and point out how to judge on the particular energy value where asymptotics has to be expected. 

Maximal strength of strong interactions means saturation of an upper bound for the total cross--section. This bound is well known after the names of Froissart,  Martin  \cite{froi,martin, martin1}. Various general and model-based arguments in favor of its saturation have been given, e.g. in \cite{mart94}.

In the QCD era the composite nature of hadrons and their finite size become important issues.
The theoretical basis of the Froissart--Martin bound has been reexamined with special attention to the assumption on the polynomial boundedness of the amplitude \cite{azimov} and the need for an additional postulate of applicability of the original approach to the hadrons composed from quarks and gluons becomes evident \cite{roy}. 

The above  saturation results from saturation of the upper limit for the $\mbox{Im}f(s,b)$. 
This limit follows from the unitarity relation for the amplitude $f(s,b)$:
\begin{equation}\label{unb}
\mbox{Im}f(s,b)[1-\mbox{Im}f(s,b)]=[\mbox{Re}f(s,b)]^2+h_{inel}(s,b),
\end{equation}
 where the inelastic overlap function $h_{inel}(s,b)$ is a nonnegative one being contribution of all intermediate inelastic channels. Then, the function 
 $\mbox{Im}f(s,b)$ should obey the following inequality  \[0\leq\mbox{Im}f(s,b)\leq 1 \] while a completely different inequality takes place for $\mbox{Re} f(s,b)$
 \[
 -\frac{1}{2}\sqrt{1-4h_{inel}(s,b)} \leq \mbox{Re} f(s,b) \leq \frac{1}{2}\sqrt{1-4h_{inel}(s,b)}.
 \]
 Saturation of the upper unitarity limit for $\mbox{Im}f(s,b)$ implies  that $\mbox{Im}f(s,b)\to 1$ at large energies and fixed impact parameters  in the region $b < r(s)$ with $r(s)$ rising logarithmically with $s$.  The two other related limits are
 $\mbox{Re}f(s,b)\to 0$ and $h_{inel}(s,b)\to 0$ at $b < r(s)$.   The above limiting values imply that at large energy 
 $\sigma_{tot}(s)\sim \ln^2 s$, $\sigma_{inel}(s)\sim \ln s$ and the ratio of real to imaginary parts of a forward scattering amplitude tends to zero
 (cf. e.g. \cite{refl}).  The linear logarithmic  increase of the $\sigma_{inel}(s)$ is a result of the self-damping of the inelastic channels \cite{bblan}.  Such self-damping results from the unitarity requirements \cite{wu} and should be taken into account under consideration of the  role of the inelastic channels due to an opening of the new ones\footnote{It should be noted that the upper bound for the inelastic cross--section obtained in \cite{wu} {\it excludes}  $\ln^2 s$-dependence for $\sigma_{inel}(s)$ at $s\to\infty$ when the ratio
 $\sigma_{tot}(s)/[(4\pi/{t_0})\ln^2(s/s_0)]$ follows its limititing behaviour, i.e. it tends to unity at $s\to\infty$.}.
 
 Thus, at large enough energies $\mbox{Im}f(s,b)$ should have  form of a smoothed step function in the impact parameter space. Such form  ultimately results in appearance of the secondary dips and bumps in $d\sigma/dt$ \cite{refl}. 
 
  \begin{figure}[hbt]
 	\vspace{-0.42cm}
 	\begin{center}
 		\resizebox{12cm}{!}{\includegraphics{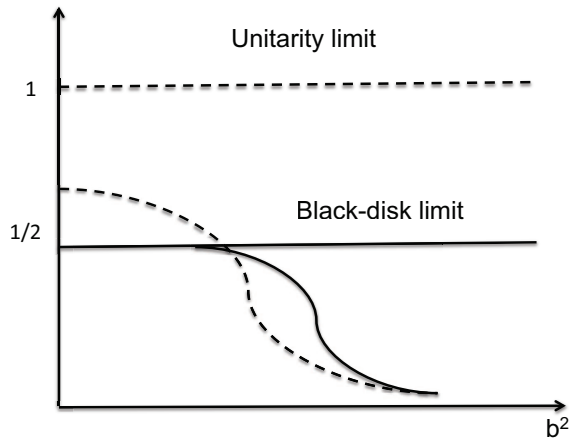}}		
 	\end{center}
 	\vspace{-2cm}
 	\caption{Schematic representation of the impact parameter dependence of the function $\mbox{Im}f(s,b)$ at the energy values in the region of $\sqrt{s}=13$ $TeV$.}	
 \end{figure}	 
 
 This conclusion also  finds its confirmation on the base of analysis of the TOTEM experimental data performed in \cite{alkin} where the two different unitarization schemes have been studied. As it was shown the eikonal unitarization leads to appearance of the sequence of the secondary dips and bumps since in this unitarization scheme the impact parameter dependence of the elastic scattering amplitude  at $\sqrt{s}=7$ $TeV$ does not leave any other possibility for interpretation aside the black disc limit saturation ($\mbox{Im} f\to 1/2$, $\mbox{Re} f\to 0$ and $h_{inel}\to 1/4$)  at low values of $b$. This and other similar to eikonal unitarization schemes do not allow crossing the black disc limit and $\mbox{Im} f(s,b)$ is forced to become similar to the step function of impact parameter (solid line at Fig.~1). Contrary, the unitarization scheme   with  black disc limit crossing \cite{refl}  (cf. Fig.~1, dashed line) does not assume appearance of the secondary dips and bumps in $d\sigma/dt$ at the LHC energy range.

 The existing accelerator experimental data are not in the asymptotical energy  region since the secondary bumps and dips have not been observed up to
 the energy $\sqrt{s}=13$ $TeV$. Even at this highest available accelerator energy value the differential cross--section has a smooth, without secondary dips and bumps, dependence on the transferred momentum in the region beyond the first dip  \cite{ken}. The data are also pointing out to excess of the black-disc limit  and not its saturation at the LHC energies. 
 It should be noted that the data obtained in cosmic ray studies do not provide an information on the differential cross--section $d\sigma/dt$.
 
 So, one may suggest to conclude on the beginning of the asymptotics in a hadron scattering by observation appearance of the secondary dips and bumps in $d\sigma/dt$. This prediction is a qualitative one, it does not rely on the consideration of any particular model for the hadron scattering and does not specify, therefore, an explicit value of the energy where the asymptotic region begins just leaving judgment on it to  the experimental studies. It is based on the impact parameter picture of hadron scattering which is more relevant for study of the asymptotic energy region than considerations of the {\it integrated} over impact parameter  quantities.
 \section*{Acknowledgements} 
 We are grateful to E. Martynov for the interesting discussions and comments.
  
\end{document}